\documentstyle[12pt]{article}
\input epsf
\begin{document}

\author{Edgardo T. Garcia-Alvarez and Fabi\'an H. Gaioli\footnote{E-mail: 
gaioli@iafe.uba.ar} \\
{\it Instituto de Astronom\'\i a y F\'\i sica del Espacio,} \\
{\it C. C. 67, Suc. 28, 1428 Buenos Aires, Argentina and }\\
{\it Departamento de F\'\i sica, Facultad de Ciencias Exactas y Naturales,}\\
{\it Universidad de Buenos Aires, 1428 Buenos Aires, Argentina}
}
\title{On the quantum electrodynamics of moving bodies}
\maketitle

\begin{abstract}
A new synthesis of the principles of relativity and quantum mechanics is
developed by replacing the Poincar\'e group for the de Sitter one. The new
relativistic quantum mechanics is an indefinite mass theory which is reduced
to the standard theory on the mass shell. The charge conjugation acquires a
geometrical meaning and the Stueckelberg interpretation for antiparticles
naturally arises in the formalism. So the idea of the Dirac sea in the
second quantized formalism proves to be superfluous. The off-shell theory is
free from ultraviolet divergences, which only appear in the process of mass
shell reduction.
\end{abstract}

\newpage

The advent of quantum theory cherished the hope of reformulating
electrodynamics free from anomalies. However, divergences were smoothed but
not completely erased by quantization. Such a disappointment was considered
as a serious trouble for the physics of that time and the progress in the
area was delayed for two decades. After the great advances achieved by the
end of the fifties, the new generation of physicists ``have learned how to
peacefully coexist with the alarming divergences of the old fashioned
theory, but these infinities are still with us, even though deeply buried in
the formalism'' \cite{Roman}. Due to this fact some workers in the field
tried to start again from the beginning formulating the so called axiomatic
quantum field theory. Their unsatisfaction was clearly summarized in the
statement of Streater and Wightman: ``...the quantum theory of fields never
reached a stage where one could say with confidence that it was free from
internal contradictions --nor the converse'' \cite{Streater}. Unfortunately
as Rohrlich \cite{Rohrlich} has pointed out, this route does not fullfil all
aspirations: ``We now have a much deeper mathematical understanding of
quantum electrodynamics, especially due to the work of axiomatic field
theorists; but we have still not solved the basic problem of formulating the
theory in a clean mathematical way, not even with all the complicated and
highly sophisticated limiting procedures presently used to justify the
results of a naive renormalization theory in simpler quantum field theories
and in lower dimensionality. The hopes and aspirations indicated in the
outlook of twenty years ago remain valid today.''

A renovating spirit was present in the more recent movement of string
theorists who decided to change some basic principles. As a consequence of
it, string models have non-local interactions which provide a way to avoid
the ultraviolet divergences from the beginning. However the price payed for
this desirable requirement is too high: we have lost the extraordinary power
of calculus and predictability of quantum field theory. This is the reason
why some theoretical physicists became conservative and, in a radical change
to the optic of the problem, tried to justify ``the unreasonable
effectiveness of quantum field theory'' \cite{Jackiw}, arguing that the
phenomenologically desirable results are provided by ultraviolet
divergences. As in the standard theoretical framework anomalies, as the
chiral one, come from the gauge non-invariance of the infinite
negative-energy sea. It is argued that ``we must assign physical reality to
this infinite negative-energy sea'' \cite{Jackiw86}. We see such
philosophical position as a new intent of rescuing the theory of the
``ether.'' Alternatively, Weinberg \cite{Weinberg} has delayed the present
difficulties for quantizing gravity reformulating the problem in this way.
He holds the point of view that the standard model and general relativity
are the leading terms in effective field theories, and so disregards the
problem of renormalizability which is only proper of a fundamental theory
still unknown (perhaps a string model).

On the contrary, the creators of the quantum field theory, such as Dirac,
held a less conservative viewpoint \cite{Salam}:

``Nowadays, most of the theoretical physicists are satisfied with this
situation, but I am not. I think that theoretical physicists have taken a
wrong way with this new facts and we would not be pleased with this
situation. We must understand that we are in front of something wrong
radically discarding the infinities from our equations; here we need to
respect the basic laws of the logics. Thinking about this point could send
us to an important advance. QED is the branch of theoretical physics about
we know more, and presumably we have to put it in order until we can make a
fundamental progress in other field theories, although this theories
continue developing under experimental basis.''

In this work we develop the foundations of a new shyntesis of the principles
of relativity and quantum mechanics. Following Dirac's advice we only
propose to reformulate QED. As our purpose is humbler than that of the
string program (conceived as the theory of everything) the change in the
basic principles is also less radical: essentially we propose to substitute
once more the standard group of external symmetries, i.e. the Poincar\'e
group for the de Sitter one. It is ironic that, approaching to the end of
this century after nine decades from Einstein did the same with the Galilei
group, we can motivate the new program rephrasing Einsten's words \cite
{Einstein}:

\smallskip\ 

It is known that Dirac's quantum electrodynamics --as usually understood at
the present time-- leads to asymmetries and inconsistencies which do not
appear to be inherent in the phenomena. Take, for example, the description
of a pair creation in an external electromagnetic field. The observable
phenomenon here always involves finite measurable quantities and does not
make any distinction between electron and positron, whereas the customary
view draws a sharp distinction between the two particles. While the electron
is interpreted as a positive energy state of the Dirac equation, the
positron is interpreted as a hole or absence of a negative energy state in
the Dirac sea.\footnote{%
The assymetry in the description is more evident from the historical point
of view. In fact the holes were originally interpreted by Dirac \cite
{Dirac30} as protons, who thought that he could explain the mass differences
by means of the interaction of the electrons of the sea.} This sea of
infinite electrons, which fills all the negative energy states of the Dirac
equation, is the responsible for ultraviolet divergences in the effective
action used for describing such phenomena.\footnote{%
This is analogous to the case of chiral anomaly discused above, and it
results specially clear from the Weisskopf derivation of the
Heisenberg-Euler Lagrangian \cite{lif}. In Sec. 2 we discuss the proper
time approach to this effective Lagrangian in which becomes clear that
divergences appear in the transition from the off-shell theory to the mass
shell.} Moreover, from the standpoint of general relativity the zero point
energy of the electromagnetic field also seems unsatisfactory since a
divergent vacuum stress tensor would imply, via the Einstein field
equations, an infinite curvature for the universe corresponding to an
infinite cosmological constant, which cannot be removed simply by performing
some sort of transfinite shift of the energy scale.

Examples of this sort, together with the unsuccessful attempts for
quantizing gravity through these methods, suggest that the phenomena of
electrodynamics as well as of gravity at a quantum level possess no
properties corresponding to the quantum field notion of the vacuum.\footnote{%
As we will see we do not discard many ``particle'' formalisms (we find more
appropriate to call them many charge formalisms) nor the notion of field. We
only attack the choice of the vacua in standard quantum field theory to
implement the charge conjugation symmetry.} They rather suggest that a
different route must be taken in order to accommodate the principles of
relativity at the quantum level. From our point of view the main difficulty
lies in the different role and interpretation of ``time'' in both theories.
In fact, while quantum mechanics privileges an absolute parameter that
labels the evolution of the system, the theory of relativity stresses the
relative character of the temporal coordinate. Therefore the first concept
of time should have the properties of a $c$-number, while the second should
be an operator due to the mixing character of the Lorentz transformations.
Thus this dual role of time poses a problem in relativistic quantum
mechanics at a first quantized level. The standard solution to this dilemma
is to give up this vessel and plunge into the sea of quantum field theory,
relegating the role of space-time coordinates to be simple parameters of the
theory. Unfortunately this mathematical artifact is achieved by means of a
choice of vacuum compatible with the idea of the Dirac sea, which actually
just swept the problem under the rug. This fact suggests us that such a dual
role of time demands the introduction of two different concepts for playing
two different roles. In other words we propose that the unification of
quantum principles with the theory of relativity requires the introduction
of an additional label to describe the events,\footnote{%
Formulations of relativistic quantum mechanics with an invariant evolution
parameter were discused in the past. According to the external group of
symmetry they can be classified as five-dimensional Galilean invariant
formulations \cite{Aghassi1,Horwitz73,Fanchi} and de Sitter ones. See
Refs. \cite{ap95a,ap95b} for a critical review about them.} increasing in
this way the dimension of the space-time manifold \cite{ap95b,ga95b,ga96a}.
We will raise this conjecture to the status of a postulate, and also
introduce another postulate, namely, laws of physics in our five-dimensional
space-time obey the principles of the special theory of relativity. These
two postulates suffice for the attainment of a simple and consistent theory
of quantum electrodynamics, based on Dirac's theory in a higher dimension.
The introduction of a ``Dirac sea'' will prove to be superfluous inasmuch as
the view here to be developed will not require ordinary time to be the
parameter which labels the quantum evolution.

\section{Kinematical Part}

Nowadays, theoretical physicists seem to be more focused on internal
symmetries than on external ones, in the search of a grand unified gauge
theory. However in the sixties a great effort was made for unifying both
symmetries, enlarging the Poincar\'e group. So for different motivations the
simplest extensions of the Poincar\'e group, such as the five-dimensional
Galilei group, the de Sitter group, and conformal group, began to be
studied, constituting the antecedents of our program.\footnote{%
In connection with this work see Refs. \cite{Castell,DeVos,Aghassi1,Aghassi2,Johnson}.%
} However the idea of enlarging the dimension of space-time to take into
account particle-antiparticle symmetries is an older fascinating idea.
Perhaps the first antecedent can be found in the works of Hinton, who built
a model of electricity associating positive and negative charges with right
and left handed helixes in higher dimensional spaces. Curiously, this
prerelativistic model developed in 1888 has an extraordinary parallelism
with the theory of Klein \cite{gard}. In Sec. 2 we discuss these ideas
through a generalization of the Schroedinger {\it Zitterbewegung} to four
dimensions \cite{bar84,ga96a}, which is related to the Stueckelberg \cite
{stu}, Wheeler and Feynman \cite{fey48,fey49,fey51,schw,nam} interpretation
of antiparticles. But in this route, the concept of time must be revisited.

Time in physics is not an {\it a priori} concept in the Newton sense, but
enters as a basic concept used to describe the laws of nature. The history
of science shows us that physics always adapts and modifies this concept in
order to simplify the laws. Then, from this point of view, there is no place
to the question why the universe has five dimensions and not four. The
important thing is that there is a set of phenomena which can be described
in a more simple and symmetrical way if we use two times instead of one. The
purpose of this work is to demonstrate that this is the case for QED.

We begin considering a five-dimensional manifold as space-time arena in
which such phenomena occur. According to the first postulate, each event in
our description has associated a point $P$ of the space-time determined by
coordinates $x^A=(x^\mu ,x^5)$ ($A=0,1,2,3,5),$ i.e. $P=P(x^A),$ which will
be called a super-event. From the second postulate the space-time is endowed
with a super-Minkowskian metric $g^{AB}={\rm diag}(+,-,-,-,-),$ so the
square of the super-arc element $dS$ reads 
\begin{equation}
dS^2=g^{AB}dx_Adx_B=g^{\mu \nu }dx_\mu dx_\nu -(dx^5)^2.  \label{ds}
\end{equation}
Any linear transformation of coordinates $x^{A^{\prime }}=L_{.B}^Ax^B+C^A$
which leaves $dS^2$ invariant will be referred to as a coordinate
transformation between two super-inertial systems. The super-Poincar\'e
group of such a transformation is the well-known inhomogeneous de Sitter
group. The other implicit assumption is that all physical laws adopt the
same form in all super-inertial frames, that is to say that they are de
Sitter covariant.

We do not analyze here all the potentialities of such a description but our
intention is to use this new framework to reformulate the physics associated
to the Poincar\'e invariance free from inconsistencies. Keeping this in
mind, let us restrict ourselves to the subset of linear transformations

\begin{equation}  \label{uno}
x^{\mu ^{\prime }}=L_{.\nu }^\mu x^\nu +C^\mu ,
\end{equation}

\begin{equation}
x^{5^{\prime }}=x^5+C^5,  \label{dos}
\end{equation}
which leaves the square of the standard arc element, $ds^2=g_{\mu \nu
}dx^\mu dx^\nu ,$ invariant, maintaining the fifth coordinate $x^5$ as a
Poincar\'e invariant parameter. This means that we are going to describe the
super-events posed in a given super-frame, forbidding boosts and rotations
between $x^5$ and any of the space-time coordinates. In this case such an
evolution parameter works as a Newtonian time in each super-frame and
introduces an absolute notion of simultaneity and retarded causality
associated to it. The fifth coordinate $x^5$ is arbitrary in principle,
however from Eq. (\ref{ds}) we see that for the particular case of motions
on the super-light cone $(dS=0)$ the coordinate $x^5$ is reduced to $s$. We
restrict our analysis of QED to this case. In Fig. 1 we show the super-light
cone and its four-dimensional projection. Note that while a super-world line
lies on the super-light cone its space-time projection lies inside the
standard light cone.
 
\medskip\ 
\begin{figure}[htb]
\centering
\epsfxsize=12truecm
\epsffile{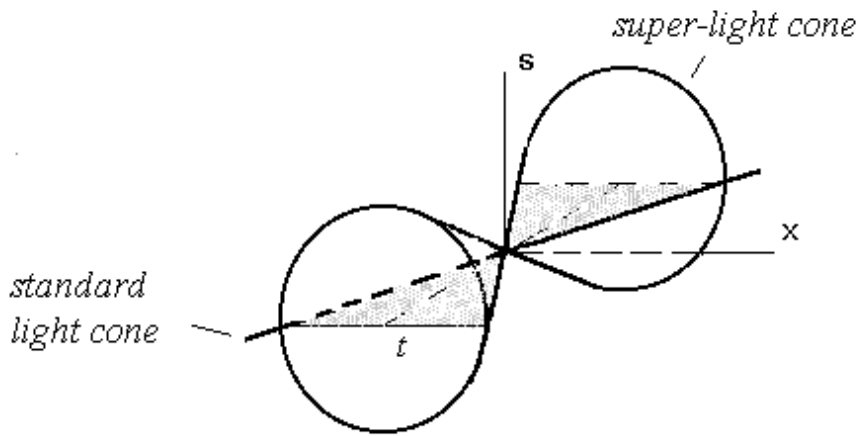}
\caption{ {\sl Super and standard light cones.}}
\end{figure}

\medskip\ 
\bigskip\

At this point one could ask what we have gained with such a description. The
immediate answer is that this description has now an invariant evolution
parameter at the classical level, preparing the land for a description at
the quantum level that avoids the lack of explicit covariance of the
standard canonical formalism. What is not so evident is that it is a natural
framework for introducing the notion of antiparticles. Moreover, as we show
in Sec. 2, the notion of retarded causality in $x^5$ for super-particles
naturally leads to the standard quantum field theoretical boundary
conditions for the Green functions on the mass-shell. That is, particles go
forward and antiparticles go backward in the coordinate time $x^0.$\footnote{%
This formalism allows us to reformulate the ``localization problem'' \cite
{Kalnay}, by following charges ``trajectories'' instead particles ones.
Moreover, the recognition that this strange notion of $x^0-$causality is the
only compatible with the requirements of relativistic quantum mechanics
enables one to eliminate Hegerfeldt's paradox \cite{Heger}.}

Let us consider the world-line of a super-event in a given super-frame. The
Poincar\'e invariance suggests us to parametrize this curve with $x^5,$ i.e.
to project the super-world-line in a hyper-plane $x^5={\rm const}$ (the
standard space-time)$.$ Thus, at any point of the projected curve (a
standard world-line), the four-velocity $\frac{dx^\mu }{dx^5}=\left( \frac{%
dx^0}{dx^5},\frac{d\overrightarrow{x}}{dx^5}\right) $ has a new key
ingredient with respect to the non-covariant description which takes the
coordinate $x^0$ as the evolution parameter, namely the rate $\frac{dx^0}{%
dx^5}.$ This new degree of freedom allows us to introduce the concept of
antiparticle just at the classical level. Generalizing Stueckelberg's ideas 
\cite{stu,fey48} we call super-particles and super-antiparticles to those
states for which $\frac{dx^0}{dx^5}$ is positive and negative respectively.
Therefore for causal propagation ($dx^5>0$), while the super-particles
propagate forward in time, the super-antiparticles propagate backward in
coordinate time. Notice that for $dx^5=0$ we cannot distinguish the two
concepts.\footnote{%
Also note that this notion is super-frame dependent, i.e. a state registered
as a super-particle from a super-inertial system can be registered as a
super-antiparticle from another super-inertial system. The same thing
happens with the notion of simultaneity associated to the coordinate $x^5,$
which looses its invariant character under the full de Sitter group
transformations.} This is the case of the photon in the standard framework,
in which we identify the fifth coordinate with the classical proper time. We
could expect that the evolution in $x^5$ also interchanges particle and
antiparticle states at a first glance. Nevertheless, as we will see below,
for the standard electromagnetic interactions this interchange is
classically forbidden and only possible at the quantum level as a
consequence of the uncertainty principle.

\section{Electrodynamical Part}

From a dynamical point of view the main difference between the Poincar\'e
and the de Sitter groups is that for the second group the operator $p_\mu
p^\mu $ is no longer a Casimir operator. The states of the new theory are
off the mass shell $p_\mu p^\mu =m^2$. They are on the super-mass shell
hyperboloid

\begin{equation}
p_Ap^A=M^2,
\end{equation}
where $M$ is a super-mass parameter. We are interested in the study of
null-super-mass states because in the classical limit they motion is
super-luminal and, as we discuss in the kinematical part, we can identify
the five coordinate $x^5$ with the proper time $s.$ So, let us begin
considering the wave equation satisfied by the non-super-massive ($M=0)$
spin-$\frac 12$ irreducible representation of the de Sitter group $\Psi $ $%
\frac {}{}$

\begin{equation}
\Gamma ^Ai\partial _A\Psi =0,  \label{irrep}
\end{equation}
where $\Gamma ^\mu =\gamma ^5\gamma ^\mu ,$ $\Gamma ^5=\gamma ^5=\gamma
^0\gamma ^1\gamma ^2\gamma ^3,$ satisfy the Dirac algebra 
\begin{equation}
\Gamma ^A\Gamma ^B+\Gamma ^B\Gamma ^A=2g^{AB}.
\end{equation}
Multiplying on the left by $\gamma ^5,$we can rewrite (\ref{irrep}) in the
Hamiltonian form

\begin{equation}
-i\frac{\partial \Psi }{\partial s}=\gamma ^\mu i\partial _\mu \Psi ,
\label{Fey}
\end{equation}
where we have identified $x^5$ with $s$ \cite{ap95b}. Eq. (\ref{Fey}) was
originally introduced by Feynman in 1948 in his dissertation at the Pocono
Conference.\footnote{%
Feynman introduced Eq. (\ref{Fey}) in a formal way and did not discuss its
geometrical meaning. He could not solve Dirac's doubts about the unitarity
of the theory either. For a nice account of these anecdotes, see the review
paper of Schweber \cite{schw}.} This is a Schroedinger equation in the
invariant parameter $s$ for the evolution of states off the mass-shell. The
mass-shell condition is satisfied by stationary states, $\Psi (x^\mu
,s)=\psi _m(x^\mu )e^{ims},$ solutions of the Dirac equation\footnote{%
The Dirac equation can be consistently introduced from first principles at a
first quantized level interpreting antiparticles as negative energy states
going backward in $x^0$-time \cite{ga95a}.}

\begin{equation}
\gamma ^\mu i\partial _\mu \psi _m=m\psi _m.  \label{Dir}
\end{equation}

The Feynman equation minimally coupled to an external electromagnetic field
is given by 
\begin{equation}
-i\frac{\partial \Psi (x,s)}{\partial s}=\gamma ^\mu (i\partial _\mu -eA_\mu
)\Psi (x,s),  \label{FeyA}
\end{equation}
where $A_\mu $ is the electromagnetic potential.

The key idea of Feynman \cite{fey51,schw} was that by Fourier transforming
in $s$ any solution $\Psi (x,s)$ of Eq. (\ref{FeyA}) a solution $\psi _m(x)$
of the corresponding Dirac equation

\begin{equation}
\left[ \gamma ^\mu (i\partial _\mu -eA_\mu )-m\right] \psi _m(x)=0
\end{equation}
can be obtained$,$ namely

\begin{equation}  \label{psi}
\psi _m(x)=\int_{-\infty }^{+\infty }\Psi (x,s)e^{-ims}ds.
\end{equation}
Hence the Fourier transform of the retarded Green function $G(x,x^{\prime
},s)$ of Eq. (\ref{FeyA})

\begin{equation}
\left[ \gamma ^\mu (i\partial _\mu -eA_\mu )-i\frac \partial {\partial
s}\right] G(x,x^{\prime },s)=\delta (x,x^{\prime })\delta (s),  \label{green}
\end{equation}
with $G(x,x^{\prime },s)=0,$ for $s\leq 0,$ enables one to derive the
corresponding mass-shell Green function $G_m(x,x^{\prime }),$ i.e.

\begin{equation}
\left[ \gamma ^\mu (i\partial _\mu -eA_\mu )-m\right] G_m(x,x^{\prime
})=\delta (x,x^{\prime }).  \label{shell}
\end{equation}
From the path integral point of view the retarded condition for the
propagator $G(x,x^{\prime },s)$ means that all the classical paths go
forward in time ($ds>0$), so the on-shell positive (negative) kinetic energy
states must go forward (backward) in coordinate time, since in the classical
limit (neglecting spin effects) we have $\frac{dx^0}{ds}=\pm \frac 1{\sqrt{%
1-v^2}}.$ This fact determines the well-known boundary conditions for $%
G_m(x,x^{\prime })$ \cite{fey49}.

Moreover if in the Fourier transformation

\begin{equation}
G_m(x,x^{\prime })=\int_0^{+\infty }G(x,x^{\prime },s)e^{-ims}ds,
\end{equation}
for the on-shell retarded Green function

\begin{equation}
G(x,x^{\prime },s)=-i\theta (s)\left\langle x\left| e^{i\gamma ^\mu \pi _\mu
s}\right| x^{\prime }\right\rangle
\end{equation}
the Schwinger formal identity

\begin{equation}
i/(a+i\epsilon )=\int_0^\infty \exp [is(a+i\epsilon )]ds  \label{sch1}
\end{equation}
is used for $a=\gamma ^\mu \pi _\mu -m$, one immediately sees that such
retarded boundary condition for $G(x,x^{\prime },s)$ naturally leads to the
Feynman $i\epsilon $ prescription for avoiding the poles in the on-shell
Green function

\[
G_m(x,x^{\prime })=\left\langle x\left| \frac 1{\gamma ^\mu \pi _\mu
-m+i\epsilon }\right| x^{\prime }\right\rangle . 
\]
This formal trick allowed Feynman to discuss external field problems of QED
keeping up at a first quantized level.

Let us go further these formal tools in order to understand the physical
grounds of them. In this formalism the state space is endowed with an
indefinite Hermitian form \cite{ap95a,ap95b}

\begin{equation}
\left\langle \Psi |\Phi \right\rangle =\int d^4x\overline{\Psi }(x)\Phi (x),
\label{pe}
\end{equation}
in which the covariant Hamiltonian or mass operator ${\cal H=}$ $\gamma ^\mu
i\partial _\mu $ is self-adjoint and the evolution operator $e^{i{\cal H}s}$
is unitary. It can be proved \cite{ga96a} that at a semiclassical level

\begin{equation}
{\rm sign}\left[ \overline{\Psi }(x,s)\Psi (x,s)\right] ={\rm sign}\frac{dx^0%
}{ds},
\end{equation}
that is super-particles and super-antiparticles states have positive and
negative norm respectively. This is the root of the indefinite character of
the ``inner product''. Frequently this fact is considered as an anomaly of
the theory, due to it is not possible to straightforward apply the standard
probabilistic interpretation. In fact this is one of the reasons why Dirac%
\footnote{%
Ironically, some years before it was Dirac himself \cite{Dirac42} who
introduced indefinite metric Hilbert spaces in quantum field theory with the
hope of removing the true anomaly: the divergences.} originally rejected the
Klein-Gordon equation. But as was shown by Feshbach and Villars \cite
{Feshbach} the indefinite metric character of the Klein-Gordon theory can be
reinterpreted in the framework of the theory of a charge. This is the
interpretation we adopt in this work.

We have defined super-particles and super-antiparticles according to the
Stueckelberg interpretation in the kinematical part. Let us now show that it
is consistent with the more familiar notion based on charge conjugation. For
making this let us note that the operation that conjugates the charge in Eq.
(\ref{FeyA}) is \cite{han,ga95b}

\begin{equation}
C\Psi (x,s)=c\Psi (x,-s),
\end{equation}
where $c=\gamma ^5K$ is the standard charge conjugation operator. The
remarkable points are that this operation coincides with the $s$-time
reversal operation in the Wigner sense \cite{ga95b}

\begin{equation}
C=S,  \label{wig}
\end{equation}
and $PcT$ looks as a ``parity'' operation in the five-dimensional space-time:

\begin{equation}
PcT=\gamma ^5Q,
\end{equation}
where

\begin{equation}
Q\Psi (x)=\Psi (-x),
\end{equation}
and $\gamma ^5$ plays the role of the ``intrinsic parity'' operator. The
identity (\ref{wig}) is the quantum analogous of a celebrated Feynman 
\cite{fey48} observation at the classical level, that charge conjugation in the
Lorentz force law is equivalent to a proper time reversal. In other words,
charge conjugation is equivalent to an inversion of the sign of $\frac{dx^0}{%
ds},$ according to the Stueckelberg interpretation for antiparticles.

In order to get a more intuitive insight about why this proper time
formalism works, let us return to the problem of particle creation in an
external electromagnetic field. In this case, the Heisenberg equations of
motion are

\begin{equation}  \label{gama}
\frac{d\gamma ^\mu }{ds}=2i\gamma ^\mu {\cal H}-2i\pi ^\mu ,
\end{equation}

\begin{equation}  \label{pi}
\frac{d\pi ^\mu }{ds}=eF^{\mu \nu }\gamma _\nu ,
\end{equation}
which form a coupled system of linear differential equations of first order
in $\gamma ^\mu $=$\frac{dx^\mu }{ds}$ and $\pi ^\mu =p^\mu -eA^\mu ,$ where
the mass operator ${\cal H}=\gamma ^\mu \pi _\mu $ is a constant of motion.

Let us restrict to the case of pure electric field, and choose the
coordinate system in such a way that $\overrightarrow{E}=E\overrightarrow{e_1%
}$, therefore the only non-vanishing components of the electromagnetic field
tensor are $F_{10}=-F_{01}=E,$ and the system of differential equations are
reduced to

\begin{equation}
\frac d{ds}\left[ 
\begin{array}{c}
\gamma ^0 \\ 
\gamma ^1 \\ 
\pi ^0 \\ 
\pi ^1
\end{array}
\right] =\left[ 
\begin{array}{cccc}
2i{\cal H} & 0 & -2i & 0 \\ 
0 & 2i{\cal H} & 0 & -2i \\ 
0 & -eE & 0 & 0 \\ 
-eE & 0 & 0 & 0
\end{array}
\right] \left[ 
\begin{array}{c}
\gamma ^0 \\ 
\gamma ^1 \\ 
\pi ^0 \\ 
\pi ^1
\end{array}
\right] ,  \label{E}
\end{equation}
plus uncoupled equations for the components $2$ and $3$ identical to the
free case \cite{bar84,ga96a}

\begin{equation}
\frac{dx^\mu }{ds}=\frac{p^\mu }{{\cal H}}+\left[ \frac{dx^\mu }{ds}(0)-%
\frac{p^\mu }{{\cal H}}\right] \cos \left( 2ps\right) -\frac 1{2p}\frac{%
d\gamma ^\mu }{ds}(0)\sin \left( 2ps\right) .  \label{dd}
\end{equation}
The system of differential equations could be exactly solved diagonalizing
the matrix of Eq. (\ref{E}). The eigenvalues are $z_{1,2,3,4}=i{\cal H}\pm 
\sqrt{-{\cal H}^2\pm 2ieE}.$ In the weak field approximation (${\cal H}^2\gg
2eE)$ the solution of this system adopts a specially simple form \cite{ga96a}
\begin{equation}
\frac{dx^0}{ds}(s)=\left. \frac{dx^0}{ds}(s)\right| _{E=0}\cosh \left( \frac{%
eE}{{\cal H}}s\right) -\left. \frac{dx^1}{ds}(s)\right| _{E=0}\sinh \left( 
\frac{eE}{{\cal H}}s\right) ,  \label{pri}
\end{equation}
\begin{equation}
\frac{dx^1}{ds}(s)=\left. \frac{dx^1}{ds}(s)\right| _{E=0}\cosh \left( \frac{%
eE}{{\cal H}}s\right) -\left. \frac{dx^0}{ds}(s)\right| _{E=0}\sinh \left( 
\frac{eE}{{\cal H}}s\right) ,  \label{seg}
\end{equation}
where $p=\sqrt{p^\mu p_\mu }$ is the free positive mass operator. The
classical picture of Eq. (\ref{dd}) together with Eq. (\ref{seg}) is a
helical motion in the space and the orbital angular momentum of this {\it %
Zitterbewegung} gives rise to the normal magnetic moment of the electron 
\cite{bar84,ga96a}. Eqs. (\ref{pri}) and (\ref{seg}) describe the classical
hyperbolic motion derived from the Lorentz force law modulated by the free 
{\it Zitterbewegung.} This quick oscillatory motion (of a Compton space-time
wavelength order) vanishes in the classical limit. Two different $s$-time
scales appear, one related to the inverse of the frequency of the {\it %
Zitterbewegung} $\frac 1{2{\cal H}}$ and the other related to the inverse of
the electric field strength $\frac{{\cal H}}{eE}.$ Then when $\frac{{\cal H}%
}{eE}\gg \frac 1{2{\cal H}},$ the {\it Zitterbewegung} does not feel the
adiabatic changes in the mean classical motion, so it works as in the free
case. The same scales also appear in the space-time trajectories. If the
minimal distance $\frac{2{\cal H}}{eE}$ between the two branches of the
hyperbola --representing particle and antiparticle solutions at the
classical level-- is greater than $\frac 1{{\cal H}},$ the particle and
antiparticle trajectories are distinguishable. However, when $\frac{2{\cal H}%
}{eE}\approx \frac 1{{\cal H}},$ such trajectories overlap, increasing the
probability that the particle jumps to the trajectory of the antiparticle
and {\it vice versa}. These jumps are reinterpreted in the standard
viewpoint --which parameterizes the dynamics with the coordinate time $x^0$%
-- as the pair creation and annihilation processes (Dirac picture\footnote{%
This picture was refined by Sauter by considering the deformation of the
energy gap produced by the electric field. Pair creation is interpreted as a
tunneling of a negative energy state (not a hole in a sea) to a positive
energy state \cite{lif}.}). Summarizing, the Schroedinger {\it %
Zitterbewegung }depicted above gives a very clear semiclassical
interpretation of such processes, which dresses the corresponding Feynman
diagrams of physical content, disregarding the concept of Dirac's sea (see
Fig. 2).

\begin{figure}[htb]
\centering
\epsfxsize=12truecm
\centerline{\epsffile{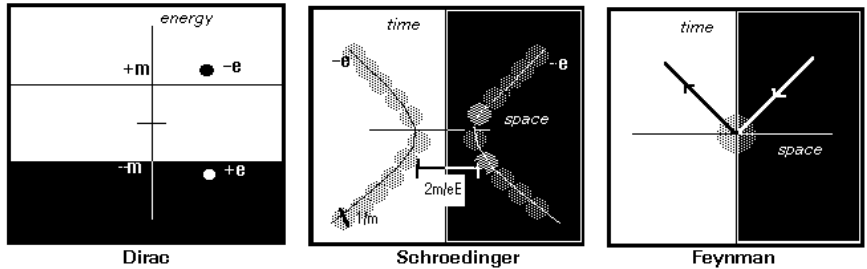}}
\caption{ {\sl Pair creation}: the dark side of relativistic quantum mechanics.}
\end{figure}

\medskip\ 
\bigskip\

At this point we disagree with some recognized field theorists that regard
Feynman's graphical method as ``a convenient pictorial device that enables
to keep track of the various terms in the matrix elements which can
rigorously derived from quantum field theory'' \cite{Sakurai}. We think that
their opinion is due to they do not completely take into account the genesis
of Feynman's ideas originally developed from the proper time method.
Unfortunately Feynman due to the misunderstanding of his dissertation at
Pocono \cite{schw} was forced to introduce his space-time visualization of
quantum electrodynamical processes in the form written in his 1949 papers 
\cite{fey49}. He relegated much of his original physical ideas and
motivations to his 1950 and 1951 papers \cite{fey51}. So there are a
generation of field theorists that have learned the derivation of Feynman
rules from Dyson's paper \cite{Dyson} rather than from Feynman's ones. In
fact when Dyson's paper appeared most of Feynman's work was still
unpublished. Unfortunately although Dyson himself remarked that ``the theory
of Feynman differs profoundly from that of Schwinger and Tomonaga,'' the
announcement of the demonstration of the equivalence (strictly speaking only
at the level of the consequences) of both theories had great impact.
Moreover the fine Schwinger calculations \cite{schw51} using a proper time
method were considered just as mathematical tools and Nambu's claims of his
deep paper of 1950 \cite{nam} 

``The space-time approach to quantum electrodynamics, as has been developed
by Feynman, seems to offer a very attractive and useful idea to this domain
of physics. His ingenious method is indeed attractive, not only because of
its intuitive procedure which enables one to picture to oneself the
complicated interactions of elementary particles, its ease and relativistic
correctness with which one can calculate the necessary matrix elements or
transition probabilities, but also because of its way of thinking which
seems somewhat strange at first look and resists our minds that are
accustomed to causal laws. According to the new standpoint, one looks upon
the world in its four-dimensional entirety. A phenomenon that will come into
play in this theatre is now laid out beforehand in full detail from
immemorial past to ultimate future and one investigates the whole of it at
glance. The time itself loses sense as the indicator of the development of
phenomena; there are particles which flow down as well as up the stream of
time; the eventual creation and annihilation of pairs that may occur now and
then, is no creation nor annihilation, but only a change of directions of
moving particles, from past to future, or from future to past; a virtual
pair, which, according to the ordinary view, is foredoomed to exist only for
a limited interval of time, may also be regarded as a single particle that
is circulating round a closed orbit in the four-dimensional theatre; a real
particle is then a particle whose orbit is not closed but reaches to
infinity ...''

\noindent
received little attention.

On the other hand most of quantum field theory treatises which intent to
incorporate the Feynman space-time visualization turn out to be
contradictory. For example they interpret field operators as operators that
create and annihilate particles in space-time points for giving an
interpretation to the Green functions. However relativistic and
non-relativistic quantum fields exhibit a striking difference concerning the
localizability of their respective field quanta \cite{Lurie}. In fact, while
in the non-relativistic case there is in principle no limitation on the
accuracy of measuring the position of a particle, the combination of
relativity and quantum theory provides an intrinsic limitation on the
measurability of the position due to the particle creation mechanism. The
understanding of such difficulties have inclined some authors to propose the
idea that Minkowsky space-time is not suitable for particle physics and its
role was essentially a historical one,\footnote{%
Although this hipothesis could work for the Poincar\'e group in the case of
free fields, strong difficulties arise at the time of introducing
interactions. Let us bear in mind that localizability and minimal coupling
are intimately linked. Moreover, this fact is not compatible with the
principle of general covariance. Notice that it would be possible to extend
this formulation to develop quantum field theory in curved space-time.}
unlike the energy-momentum space which would be fundamental \cite{Bacry}. On
the contrary, in our proposal we prefer to leave Poincar\'e group and retain
the localizability in Minkowsky space-time.

Summarizing, those field theories which desire to keep Feynman diagrams
interpretative picture, must give up the Poincar\'e group. There is no
space-time localization of particles in this framework. There is only space
time localization of charges off the mass-shell.

In order to reinforce our pictorial image of the Fig. 2 let us derive the
one-loop effective action $W^{(1)},$ which describes the pair creation in an
external electromagnetic field, from an argument purely based on the proper
time formalism. As $W^{(1)}$ is $i$ times the closed loop amplitude $L$, let
us compute $L$ using the proper time formalism. First, let us evaluate the
amplitude for a super-particle at $x^\mu $ and polarization $k$ at time $s=0$
remains in the same point and with the same polarization at time $s.$ As a
consequence of the indefinite metric ($\ref{pe})$, the spectral resolution
of the identity is

\begin{equation}
I=\int d^4x\sum_{jk}\gamma _{jk}^0\left| j,x^\mu \right\rangle \left\langle
k,x^\mu \right| .
\end{equation}
Then the expression of such an amplitude per unit of proper time for all the
degrees of polarization is $\frac 1s\sum_{jk}\gamma _{jk}^0\left\langle
k,x^\mu \left| e^{i(\gamma ^\mu \pi _\mu )s}\right| j,x^\mu \right\rangle .$
The above process is represented through an open diagram in the
five-dimensional space-time, but it is a closed loop in four dimensions \cite
{dav55}. Restricting the formalism to the mass-shell by means of a Fourier
transformation in proper time with the causal prescription and summing the
contributions of each space-time point, we finally have

\begin{equation}
W^{(1)}=i\int \int_0^\infty \frac 1s\sum_{jk}\gamma _{jk}^0\left\langle
k,x^\mu \left| e^{i(\gamma ^\mu \pi _\mu )s}\right| j,x^\mu \right\rangle
e^{-ims}dsd^4x.  \label{wef}
\end{equation}
Schwinger, using quantum field theory, obtained Eq. (\ref{wef}), which
became the starting point of his 1951 seminal paper \cite{schw51,fey51}.

The procedure used in the calculation of $W^{(1)}$ also shows that the
ultraviolet divergences only appear after the reduction of the off-shell
amplitude on the mass shell. Note that this circumstance also suggests a
natural regularization method based on a small mass dispersion \cite{fey51}.
Our alternative explanation does not involve the infinite amount of energy
and charge of the Dirac sea in order to consider antiparticles, and in this
way it avoids the infinities introduced in the standard theory from the very
beginning. This is the reason why closed loops do not appear in the
off-shell theory.

Until now we have only discussed the theory of external fields. In order to
concluding, let us briefly discuss the radiative process.

Using this formalism and his operator calculus, Feynman presented at Pocono
a closed expression for a system of spin half charges interacting via the
quantized electromagnetic field for the case in which only virtual photons
are present. In the particular case of one charge it reads \cite{fey51,schw} 
\begin{eqnarray}
\Psi (x,s) &=&\exp \left\{ -i\left[ \int_0^s\gamma ^\mu (s^{\prime })\pi _\mu
(s^{\prime })ds^{\prime }\right. \right.  \label{pocono} \\
&&\left. \left. +e^2\int_0^s\int_0^s\gamma ^\mu (s^{\prime })\gamma _\mu
(s^{\prime \prime })\delta _{+}\{[x_\mu (s^{\prime })-x_\mu (s^{\prime
\prime })]^2\}ds^{\prime }ds^{\prime \prime }\right] \right\} \Psi (x,0), 
\nonumber
\end{eqnarray}
where $\delta _{+}\{[x_\mu (s^{\prime })-x_\mu (s^{\prime \prime })]^2\}$ is
the Green function of the d'Alembertian with Feynman's boundary conditions.
From the second term of Eq. (\ref{pocono}) Feynman showed that the radiative
corrections of QED can be derived. The analogy between the phase of Eq. (\ref
{pocono}) and the Wheeler-Feynman action \cite{Wheeler,fey48} for classical
electrodynamics is remarkable. In fact the only substantial difference is
the boundary conditions (half-advanced and half-retarded) chosen for the
d'Alembertian Green function. The right boundary conditions for QED can be
obtained from the retarded condition of the off-shell theory. This fact
strongly suggests that Eq. (\ref{pocono}) could be derived, from first
principles, from a de Sitter invariant formulation of QED.

For one super-particle (antiparticle) the de Sitter invariant equations read

\begin{equation}  \label{ba}
\Gamma ^A(i\partial _A-eA_A)\Psi =0,
\end{equation}

\begin{equation}
\partial _AF^{AB}=e\overline{\Psi }\Gamma ^B\Psi ,  \label{bb}
\end{equation}
where the super-potential $A^A$ =($A^\mu ,A^5)$ arises from a natural
extension of the gauge principle \cite{Horwitz}. The standard four-potential
can be obtained from $A^A$ integrating the first four components in the
proper time

\[
A^\mu (x^\nu )=\int_{-\infty }^{+\infty }A^\mu (x^\nu ,s)ds,
\]
as in the case of the matter fields. (The exponential factor does not appear
in this case because the photon is non-massive. Note also that the
transformation $A^\mu (x^\nu ,s)\rightarrow $ $A^\mu (x^\nu ,-s),$ ($%
ds\rightarrow -ds),$ leads to the standard notion of charge conjugation for
the potentials.) 

\section*{Note added in proof}

After completing this work we discovered a review paper of Fanchi \cite{nap1} 
and the closely related works of Herdegen \cite{nap2} and Kubo \cite{nap3}.

\section*{Acknowledgments}

We are grateful to the organizers of the First International Colloquium on 
`Actual Problems in Quantum Mechanics, Cosmology, and the Primordial Universe' 
and the `Foyer d'Humanisme' for thier warm hospitality. 

We want to express our acknowledgment to Juan
Le\'on Garc\'\i a, Lawrence Horwitz, and Marek Czachor for their
encouragement.

\end{document}